\documentclass[ag,natbib]{copernicus}

\usepackage{color}
\usepackage{graphicx}
\usepackage{natbib}
\usepackage{mathptmx}      
\usepackage{amsmath}
\usepackage{amsfonts}
\usepackage{amssymb}
\usepackage{amsbsy}

\usepackage{latexsym}
\begin{document}

\title{{Relativistic transformation of phase-space distributions}}

\author[1,2]{{R. A. Treumann}\thanks{Visiting the International Space Science Institute, Bern, Switzerland}}
\author[3]{{R. Nakamura}}
\author[3]{{W. Baumjohann}}

\affil[1]{Department of Geophysics and Environmental Sciences, Munich University, Munich, Germany}
\affil[2]{Department of Physics and Astronomy, Dartmouth College, Hanover NH 03755, USA}
\affil[3]{Space Research Institute, Austrian Academy of Sciences, Graz, Austria}

\runningtitle{Relativistic phase-space distributions}

\runningauthor{R. A. Treumann, R. Nakamura, and W. Baumjohann}

\correspondence{R. A.Treumann\\ (rudolf.treumann@geophysik.uni-muenchen.de)}

\received{ }
\revised{ }
\accepted{ }
\published{ }


\firstpage{1}

\maketitle

\begin{abstract}
We investigate the transformation of the distribution function in the relativistic case, a problem of interest in plasma when particles with high (relativistic) velocities come into play as for instance in radiation belt physics, in the electron-cyclotron maser radiation theory, in the vicinity of high-Mach number shocks where particles are accelerated to high speeds, and generally in solar and astrophysical plasmas. We show that the phase-space volume element is a Lorentz constant and construct the general particle distribution function from first principles. Application to thermal equilibrium lets us derive a modified version of the {isotropic} relativistic thermal distribution, the modified J\"uttner distribution {corrected for the Lorentz-invariant phase-space volume element}. Finally, we discuss the relativistic modification of a number of plasma parameters.

\end{abstract}

\section{{Introduction}}
\noindent One-particle distribution functions are defined as the probability of finding a certain number $d\,{\cal N}_\alpha$ of particles of species $\alpha$ at time $t$ in a given phase-space volume element $dx^3dp^3$ as
\begin{equation}
f_\alpha(\mathbf{x},\mathbf{p},t)=\frac{d\,{\cal N}_\alpha}{dx^3dp^3}
\end{equation}
Since the above function is a probability function this expression can also be read as a normalisation condition 
\begin{equation}
dx^3dp^3\frac{f_\alpha(\mathbf{x},\mathbf{p},t)}{d\,{\cal N}_\alpha} =1
\end{equation}
saying that the product of the probability divided by the number density of particles in the phase space volume under consideration must be one, thus being invariant. Under special-relativistic conditions the definition of such a distribution poses a non-trivial problem. In this Letter we define such a distribution from first principles showing that of the critical points the first is to confirm that the phase-space volume element must correctly be made Lorentz invariant. Once this has been done, it is possible to define a distribution function valid for the entire phase space. 

\section{Construction of phase-space distributions}
Clearly, if the distribution function is a proper probability, it is by its nature invariant with respect to any coordinate transformation of the spatial and momentum space volume elements, if the number of particles in the volume element is kept constant. This follows from Liouville's theorem and the dynamics of the particles (at least as long as no particles are lost or added to the volume element under consideration). However, a proper transformation of the volume element might be in place. In the Galileian non-relativistic case there is no problem at all. Problems arise when the particle dynamics becomes relativistic. While the functional form of the phase-space probability in such transformations is conserved, its dependence on the new phase-space coordinates $\mathbf{x}',\mathbf{p}'$ will change during the transformation. This has been noticed long ago \citep[for the basic arguments see, e.g.,][Section 10]{landau1975}, and the transformation of the phase-space volume becomes non-trivial. In fact, in order to arrive at a consistent formalism and understanding one cannot simply transform the above formula; rather one must return to the exact definition of the distribution function given by
\begin{equation}
F(\mathbf{x},\mathbf{p})= \frac{1}{{\cal N}}\prod\limits_i\delta\big[\mathbf{x}-\mathbf{x}_i(t)\big]\delta\big[\mathbf{p}-\mathbf{p}_i(t)\big]
\end{equation}
where $\mathbf{x}_i(t),\mathbf{p}_i(t)$ are the dynamical coordinates of the $i$th particle. Since $F$ sums up all points in phase space which are occupied by the ${\cal N}$ particles numbered $i=1,\dots,{\cal N}$, it is exact. Accordingly, the exact distribution of one single particle is defined as
\begin{equation}\label{eq-1}
F_j(\mathbf{x},\mathbf{p})= \delta\big[\mathbf{x}-\mathbf{x}_j(t)\big]\delta\big[\mathbf{p}-\mathbf{p}_j(t)\big]
\end{equation}
and the exact distribution of a group $\delta{\cal N}$ of particles is given by
\begin{equation}\label{eq-2}
\delta F_{i\dots j}(\mathbf{x},\mathbf{p})= \frac{1}{\delta{\cal N}_{i\dots j}}\prod\limits_{\ell=i}^{\ell=j}\delta\big[\mathbf{x}-\mathbf{x}_\ell(t)\big]\delta\big[\mathbf{p}-\mathbf{p}_\ell(t)\big]
\end{equation}
This all looks simple and is simple as long as one does not include relativity. Then, however, particles are all in motion, each in its own proper frame and with its own proper direction of velocity and momentum. The transformation from that frame into another frame, i.e. into the observer's stationary frame involves the Lorentz transformation of each particle's frame separately. In the (primed) particle's proper frame $(\mathbf{x}',\mathbf{p}')$ the particle is located at the origin and the proper-frame distribution $F^0_j$ Eq. (\ref{eq-1}) reads
\begin{equation}\label{eq-3}
F_j^0(\mathbf{x}',\mathbf{p}')= \delta(\mathbf{x}')\delta(\mathbf{p}')
\end{equation}
while Eq. (\ref{eq-1}) is the expression in another frame with respect to which the particle is displaced, has momentum $\mathbf{p}_j(t)$ and moves with relative velocity $\mathbf{v}_j(\mathbf{x}_j,t)$ with respect to the new frame. Taking the axes of the two frames all parallel to each other and assuming that the particle instantaneously moves along $x'$, and defining \boldmath ${\beta}$\unboldmath $=\mathbf{v}/c, \gamma=1/\sqrt{1-\beta^2}=\sqrt{1+p^2/m^2c^2}$ with $m$ particle mass, momentum $\mathbf{p}=mc\gamma\,\mathbf{\beta}$, thus $\mathbf{\beta}=\gamma^{-1}\sqrt{\gamma^2-1}$, and time $t\to ct$ one has as usually for the spatial coordinates
\begin{equation}\label{eq-XLT}
[\,x',\ y',\ z',\ t'\,]=\big[\,\gamma(p)\,(x-\beta t),\ y, \ z, \ \gamma(p)\,(t-\beta x)\,\big]
\end{equation}
Inserting into Eq. (\ref{eq-1}) then yields for the transformed exact single-particle distribution function
\begin{eqnarray}
\hspace{1.1cm}F_j(\mathbf{x},\mathbf{p},t)&=& \delta\Big\{\mathbf{x}-\big[\,\gamma_j(p)\,(x-\beta_j t),\ y, \ z\,\big]\Big\}\times\nonumber\\
&\times&\delta\big[\mathbf{p}-\mathbf{p}'_j(t)\big]\delta\big[t-\ \gamma_j(p)\,(t-\beta_jx)\big]
\end{eqnarray}
In this expression $\mathbf{p}'_j=\mathbf{p}_j$ because there is only one (instanteneous) relative velocity between the moving $j$th particle's and the observer's frame. However, $\mathbf{\beta}$ is a function of $\gamma(p)$, and though the functional form of the probability distribution (represented by the product of $\delta$-functions) is conserved, the dependence on phase-space coordinates becomes complicated.

In considering many particles with their different proper velocities, proper momenta and different directions of velocities, one needs to rotate the axes of all the proper frames in performing the various Lorentz transformations. This inhibits the use of the exact equation Eq. (\ref{eq-2}) even in the case of a completely isotropic angular distribution of the various velocities. Moreover, stationarity of the distribution becomes illusionary for several reasons, the simplest one that it is impossible to assign initial conditions to all particles at a fixed time $t=0$ for the obvious reason that this requires time. This time is needed for jumping from one proper frame to the next and, thus, is limited by relativity and the requirements of relativistic synchronisation. It can be achieved only for a small limited number of particles.

\section{Lorentz invariance of full phase-space volume element}
The only way out of this dilemma is to neglect the proper particle motions altogether and assume that by some mysterious (thermodynamic) reason the particles have managed to reach a stationary final state where their momenta are statistically distributed. In this case a common phase-space volume can be attributed to them all, and each particle moves with its (thermal) momentum with respect to this \textit{common proper} frame. If the momenta are isotropically distributed in space then the probability distribution will not depend on the direction of velocities. It can then be considered as averaged over all angles and depends just on the modulus of the momentum $p$. 

The way to do this is to take one single particle and to smooth it out so as to fill a certain infinitesimal but finite proper volume $d^3x'_jd^3p'_j$, which is actually a volume element, while defining a probability of finding the particle in this volume $f_j(\mathbf{x}',\mathbf{p}')$ which now is a continuous function which vanishes outside the volume element and is normalised according to
\begin{equation}
1=\int\limits_{{\cal V}_j} f_j(\mathbf{x}'_j,\mathbf{p}'_j)d'^3x_jd'^3p_j
\end{equation}
with integration just over the infinitesimal $j$th volume. It is unity because the infinitesimal volume contains only the one $j$th particle, and $f_j=0$ outside it. In principle the latter fact would allow for extending the integration over the entire phase space volume.

Within this small volume element it can be said that prescribing the initial state has been done at a fixed time $t'=0$. This is correct for sufficiently small volumes such that synchronism is warranted. It thus becomes possible to Lorentz transform the infinitesimal volume to the un-primed observer frame. Since the particle velocity at fixed time is linear, one may rotate both frames to coincide in the directions of their axes. 

Applying the spatial transformation Eq. (\ref{eq-XLT}) at $t=0$ and observing that only the coordinate $x'$ is affected by the Lorentz transformation becoming contracted when transforming to the un-primed observer frame, one immediately finds \citep[cf., also,][his Eqs. (2.10.11) \& (2.10.12) for the transformation of the particle 4-current density]{weinberg1973} that
\begin{equation}\label{eq-spatial}
d^3x_j'=\gamma_j(p)d^3x_j
\end{equation}
where the index $j$ at $d^3x_j$ reminds one that the observer's frame has been rotated into the particle frame, and $\mathbf{p}_j$ is the particle momentum of the smeared out infinitesimal volume of the particle in the observer's frame. In the proper frame of the particle the momentum has zero average $\langle \mathbf{p}'_j\rangle=0$. In the frame of the observer the finite infinitesimal momentum space volume $d^3p_j$ is located at the tip of the particular particle-momentum vector $\mathbf{p}_j$. The proper Lorentz factor $\gamma_j(p)$ depends only on the modulus of $\mathbf{p}_j$.

One may also understand the small infinitesimal momentum space volume as the \textit{thermal uncertainty} of the particle momentum that is caused by the quiver velocity of the particle. This quiver velocity is the rms velocity $\langle{u}\rangle_j=\sqrt{(\Delta\mathbf{u}_j)^2}$ of the particle's proper motion, its thermal speed. It may be assumed to be isotropic though this is not necessary if this thermal speed is anisotropic due to some other reason. 

With all this in mind we can proceed to the transformation of the momentum space element. This is more involved than the transformation of the spatial volume. It requires returning to the definition of the momentum $\mathbf{p}_j=mc\gamma_j(p)\beta_j$ and taking into account that the particle energy and momenta in relativity are not independent, being related via the relation $pc={\cal E}$ or ${\cal E}=m\gamma(p) c^2$. This implies that the transformation of momentum and energy has to be done together. This becomes
\begin{equation}
\left[
\begin{array}{c}
  {\cal E}'    \\
  cp'_x   
\end{array}
\right] = \gamma(p)
\left[
\begin{array}{cc}
  1& -\beta     \\
  -\beta &  1 
\end{array}
\right]\left[
\begin{array}{c}
  {\cal E}   \\
  cp_x   
\end{array}\right], \qquad 
\begin{array}{rcl}
  p'_y&=& p_y     \\
  p'_z& =&p_z 
\end{array}
\end{equation}
The variations in energy and momentum become
\begin{eqnarray}
\hspace{1.5cm}dp'_x&=&\frac{\gamma(p)}{c}\big[\,c\,dp_x-\beta d{\cal E}\big] \\
d{\cal E}'&=&-\gamma(p)\big[c\beta\,dp_x-d{\cal E}\big] =0
\end{eqnarray}
The last expression follows because, in the particle rest frame, the energy remains unaffected by the transformation, yielding as usually $d{\cal E}=c\beta\,dp_x$. Hence,
\begin{equation}
dp'_x=\gamma(p)\,dp_x\big(1-\beta^2\big)=\frac{dp_x}{\gamma(p)}
\end{equation}
which, of course, is just a special case of the well-known transformation law
\begin{equation}
\frac{dp'_x}{mc\gamma(p')}=\frac{dp_x}{mc\gamma(p)}, \quad\mathrm{with}\quad\gamma(p')=1
\end{equation}
and the single-particle phase-space volume including Eq. (\ref{eq-spatial}) is an invariant under Lorentz transformation
\begin{equation}
d^3x'_jd^3p'_j=d^3x_jd^3p_j
\end{equation}
This holds for each particle separately such that one can define a single-particle distribution function in the observer's frame by
\begin{equation}
f_j(\mathbf{x}_j,\mathbf{p}_j)d^3x_jd^3p_j=f_j (\mathbf{x}'_j,\mathbf{p}'_j)d^3x'_jd^3p'_j
\end{equation}

\section{Global phase-space distribution}
One would wish to add up all single particle distributions to one distribution which covers the entire phase space. This task encounters the problem that so far the two frames, the $j$th particle frame and the observer frame, have been aligned. In order to obtain a similar result this must be done for all particles. Hence, simply adding the distribution functions is inappropriate. One needs to introduce a distribution of angles for the various particle moments as seen from the stationary  observer frame. This can be achieved with the help of the solid angle $\Omega_p$ and the distribution $\Phi(\Omega_p)$. 
Then in the stationary frame one defines the probability of finding all ${\cal N}$ particles in the volume of phase space occupied by all the smeared-out infinitesimal though finite volumes of the particles (assuming that they do not overlap, i.e. in the absence of collisions) 
\begin{eqnarray}
f(\mathbf{x},\mathbf{p})\!\!\!\!\!\!\!\!\!&&\!\!\!\!\!\!\!\!\! d^3\!\!x\ d^3\!\!pd\Omega_p= \prod\limits_{j=1}^{\cal N}f_j(\mathbf{x}'_j,\mathbf{p}'_j)\Phi(\Omega_p)\ d\Omega_p \ d'^3\!\!x{\!_j}\ d^3\!\!p'_{\!j} \cr
&=&\prod\limits_{j=1}^{\cal N}f_j[\mathbf{x}'_j(\mathbf{x},\mathbf{p}),\mathbf{p}'_j(\mathbf{p},{\cal E})]\Phi(\Omega_p)\ d\Omega_p \ d^3\!\!x\ d^3\!\!p
\end{eqnarray}
For an isotropic distribution of particle momenta the integral over the solid angle gives just a factor $4\pi$. This yields for the probability distribution in the stationary frame the final result
\begin{equation}
f(\mathbf{x},\mathbf{p},\Omega_p)\ d\Omega_p=\prod\limits_{j=1}^{\cal N}f_j[\mathbf{x}'_j(\mathbf{x},\mathbf{p}),\mathbf{p}'_j(\mathbf{p},{\cal E})]\Phi(\Omega_p)\ d\Omega_p
\end{equation}
Normalisation of this distribution must as usually be done to the total number ${\cal N}$ of particles respectively the average particle density $N={\cal N}/{\cal V}$. With this definition, the relativistic particle phase-space distribution can indeed be written as a single average distribution while the directions of the different single particle velocities and thus the various rotations of the ${\cal N}$ proper coordinate frames is taken care in the angular dependence of $f (\mathbf{x},\mathbf{p},\Omega_p)$, which eliminates the otherwise required dependence of the volume element on the proper particle-speed direction as given by \citet[][and references therein]{debbasch2001}. 

Clearly, this angular distribution must be introduced by hand, making a reasonable guess. This, however does not provide any problem as it can be chosen to be inherent to the initial model. Again, in the case of an isotropic angular distribution it just gives, after integration, a factor of $4\pi$.

\section{Applications}
The evolution of the relativistic distribution function $f(\mathbf{x},\mathbf{p},\Omega_p)$ in a dynamical system is subject to the relativistic Vlasov equation
\begin{equation}
\frac{\partial f}{\partial t}+[\!\![{\cal H},f]\!\!]=0, \quad (\mathrm{with}\ {\cal H}\ \mathrm{the \ Hamiltonian})
\end{equation}
written in the 3+1-split non-covariant representation, where $[\!\![\dots]\!\!]$ is the classical Poisson bracket. In fact, here ${\cal H}(\mathbf{x,p})$ is the 1-particle Hamiltonian. Under equilibrium conditions the time dependence is dropped, and in thermal equilibrium the left-hand side is identically zero. 

\paragraph*{Relativistic thermal equilibrium distribution}
The stationary thermal equilibrium distribution assumes the form of a relativistic variant of the Maxwell-Boltzmann distribution. Its determination poses a nontrivial problem \citep[cf., e.g.,][and references therein]{dunkel2007} as the thermal equilibrium under relativistic conditions is defined more subtly, yielding different forms of the so-called Maxwell-J\"uttner distribution. Its functional form must in fact be derived from putting the Boltzmann collision term 
\begin{eqnarray}
\hspace{0.5cm}0&=&\int d^3p'\, \sigma_\mathit{coll}\ d\Omega_\mathit{coll}\ \left[(f'f)^*-f'f\right]\ \times \cr
&\times&\sqrt{(\mathbf{p'\cdot p})^2+m^4c^4\left[\gamma^2(p')\gamma^2(p)-1\right]}/mc\gamma\,(p')
\end{eqnarray}
to zero which, in the above equation, has been taken to vanish identically. Here $\sigma_\mathit{coll}$ is the collisional cross section, and $d\Omega_\mathit{coll}$ the solid collisional deflection angle, and primed and un-primed quantities refer to the two colliding particles, while the asterisk refers to quantities after collision. This then yields $[(f'f)^*-f'f]=0$ and, ultimately, the famous exponential Boltzmann-Gibbs dependence on energy:
\begin{equation}\label{reldist}
f= \Lambda\ \exp\ \left[-\Theta_0\,mc^2\gamma\,(p)-\sum\limits_{i=1}^3\Theta_i cp^i\right]
\end{equation}
The Lagrangean multipliers $\Lambda, \Theta_0, \Theta_i$ must be determined as usual from particle number and global energy-momentum conservation; the 4-vector components $\Theta_\nu$ have dimension of inverse energy which suggests that they relate to a particular inverse-temperature 4-vector. Consistency with the collisionless part of the Vlasov equation yields that $\Lambda=$\,const does not depend on space, while $\Theta_\nu$ could become a linear 4-vector function of space. Under isotropic conditions, however,  only the average energy comes into play, and one may put all $\Theta_i=0, i=1,2,3$, keeping only the constant $\Theta_0\neq 0$. 

\vspace{2mm}\noindent
\emph{General form of distribution.} This approach then leads to the canonical form of the isotropic relativistic distribution function
\begin{equation}\label{reldist-1}
f_\mathit{rel}[\gamma(p)]= \Lambda\ \exp\ \left[-\Theta_0\,mc^2\gamma\,(p)\right]
\end{equation}
with the only remaining problem to determine the two Lagrangean multipliers $\Lambda$ and $\Theta_0$. The latter can simply be guessed from the requirement that the argument of the exponential should be dimensionless, which yields $\Theta_0=1/T$ \citep[for rigorous argument, cf.][]{israel1963} identifying it with the inverse of temperature $T$ of the macroscopic system of particles. Determination of $\Lambda$ requires solving for the normalisation of the distribution. Here the different approaches diverge, and there has no unambiguous consensus be found. 

It is argued \citep{dunkel2007} that, if rigorous Lorentz invariance is imposed (as has been used in our strict definition of the general non-stationary distribution function\footnote{It is important to note that this is the only case which accounts for the required general Lorentz invariance of the complete phase-space volume element, and only under the condition of its validity is it possible to define a global phase-space distribution function.}) the correct (non-angular-dependent part of the) relativistic thermal-equilibrium distribution should become the \textit{modified-}J\"uttner distribution. \citep[The ordinary Maxwell-J\"uttner distribution function was derived by F.][who obtained it imposing translational invariance in momentum space only.]{juttner1911}

\citet{dunkel2007} chose a general approach starting from the maximum-entropy principle for the Boltzmann-Shannon (relative) entropy density (actually the entropy density per particle)  $s_0$ as a functional of the distribution function
\begin{equation}
s_0[f(\mathbf{p}|\ell(\mathbf{p})]=-\int_{{\cal V}_p} d^d\mathbf{p}\ f(\mathbf{p})\log\,\left[(mc)^d f(\mathbf{p})\right]
\end{equation}
where $\ell(\mathbf{p})=(mc)^{-d}$ is the constant momentum space density needed for normalising the argument of the logarithm, and ${\cal V}_p$ the momentum space volume. This is subject to the constraints
\begin{equation}
1=\int_{{\cal V}_p}d^d\mathbf{p}\,f(\mathbf{p}),\quad {\cal E}=\int_{{\cal V}_p}d^d\,f(\mathbf{p})\epsilon(\mathbf{p})
\end{equation}
where ${\cal E}$ is the average energy density, i.e. the scalar pressure which is the trace of the energy-momentum tensor $T^{\mu\nu}$,  and $\epsilon(\mathbf{p})=mc^2\gamma(\mathbf{p})$ the relativistic particle energy. Defining the Haar measure to momentum space, they propose the modified-J\"uttner distribution for arbitrary dimensions $d$. Unfortunately, however, Lorentz invariance is applied only to momentum space. 

On using the Boltzmann collision-integral approach, \citet[][and references therein]{dagdug2010} obtain another form in arbitrary dimensions $d$ which, for $d=3$, reduces to the original Maxwell-J\"uttner function \citep[already given earlier by][]{israel1963}. 

\vspace{2mm}
\noindent\emph{J\"uttner-like thermal distribution.} However, all these approaches only consider the momentum-space part of the volume element neglecting the transformation of the spatial coordinates, thus keeping the factor $p^0=mc\gamma(p)$ in the denominator of the momentum-space volume element. This is either not justified at all or it is argued that the particles are all confined to a fixed box which is unaffected by the Lorentz transformation and invariance. However, the momentum and configuration space volume elements the product of which forms the phase-space volume element, are not independent, as we have demonstrated above. Even in this case of a fixed outer box, the particle's proper spaces experience linear Lorentz contractions when seen from the stationary frame of the observer, i.e. from the box-frame perspective. The consequence is that the extra proper Lorentz factor $\gamma(p)$ in the phase-space volume element cancels thereby guaranteeing and restoring Lorentz invariance. The correct form of the normalisation condition is thus
\begin{equation}
N^\nu\!\!=\!\!\int d^3\!\!\!p\,p^\nu\!\!f_\mathit{rel}[\gamma(p)], \quad T^{\mu\nu}\!\!=\!\!\int d^3\!\!\!p p^\mu\!\!p^\nu\!\!f_\mathit{rel}[\gamma(p)]
\end{equation}
where $N^\nu$ is the particle-current density 4-vector, and $T^{\mu\nu}$ the average energy-momentum tensor. When taking into account the functional dependence of the relativistic distribution Eq. (\ref{reldist-1}) on $\Theta$, both can be derived from the Integral
\begin{equation}
{\cal I}=\int d^3pf_\mathit{rel}[\gamma(p)]
\end{equation}
by differentiation with respect to $\Theta$. For instance, with $p^0=mc\gamma$, one has for the average particle density in the observer frame (the stationary box)
\begin{equation}
N\equiv N^0=-\frac{\Lambda}{c}\frac{\partial{\cal I}}{\partial \Theta_0}
\end{equation}
Calculation of the integral is straightforward. Following earlier work \citep{juttner1911,israel1963,dagdug2010} one introduces spherical-harmonic coordinates in momentum space and uses the usual representation of the Lorentz factor by hyperbolic functions as
\begin{eqnarray}
\hspace{0.5cm}p^0&=&mc\ \cosh\xi \cr
p^i&=&mc\ \sinh\xi\ \left(\sin\theta\cos\phi,\sin\theta\sin\phi,\cos\theta\right)
\end{eqnarray}
with $i=1,2,3$ and $0< \xi<\infty$. This yields
\begin{eqnarray}
\hspace{5mm}{\cal I}&=&\int d^3\!\!\!p\ \mathrm{e}^{-\Theta_0\, mc^2\cosh\xi} \cr
&=&4\pi(mc)^3\!\!\!\int d\xi\sinh^2\xi\cosh\xi\mathrm{e}^{-\Theta_0\, mc^2\cosh\xi}
\end{eqnarray}
If one choses $x=\cosh\xi$ as dummy integration variable, the boundaries of the integral become $1\leq x<\infty$. Performing the integration yields
\begin{equation}\label{theta-1}
{\cal I}=\ 4\pi(mc)^3 \frac{K_2(mc^2\Theta_0)}{mc^2\Theta_0}, \qquad \Theta_0=\frac{1}{T}
\end{equation}
an expression in which the second order modified Bessel function $K_2(z)$ appears instead of $K_1(z)$. This result is different from those obtained without taking into account the correct Lorentz invariance of the full phase-space volume element, including the configuration space. 

Taking the derivative as prescribed above we find
\begin{equation}\label{lambda}
\Lambda=\frac{N^0}{4\pi m^2T^2}\left[3K_2\left(\frac{mc^2}{T}\right)+\frac{mc^2}{T}K_1\left(\frac{mc^2}{T}\right)\right]^{-1}
\end{equation}
This is slightly more complicated than J\"uttner's normalisation but takes into account the correct Lorentz invariance of the phase-space volume element as required by relativity. In any application it is therefore preferable over the former. Eqs. (\ref{reldist-1}), (\ref{lambda}), and (\ref{theta-1}) define the correct relativistic isotropic thermal equilibrium distribution. 

{Like the conventional J\"uttner distribution, the modified J\"uttner distribution given here refers to thermal equilibria being just the relativistic version of the Boltzmann energy distribution when taking into account Lorentz invariance of the phase-space volume element. Observations of high energy particles, for instance Cosmic Ray spectra or spectra of energetic particles in the solar wind and in the vicinity of shocks, frequently deviate quite strongly from such distributions exhibiting long non-thermal power law tails in energy and velocity. The physics of generation of such tails is still a problem of intense research. Clearly such distributions are not described by thermal equilibria; it is usually assumed that highly non-linear wave particle interactions in non-equilibrium systems, i.e. mostly in the presence of collisionless shock waves, or else plasma turbulence, would be capable of generating particle spectra and distributions with extended tails. Since in most cases those tails reach far into the relativistic domain one would from the beginning assume that the non-linear problem of tail generation should be treated relativistically working with  relativistic thermal distributions as initial equilibrium conditions which, when exposed to turbulence or non-linear processes, finally evolve relativistic tails. These questions are particularly important in astrophysical applications to energetic particle spectra.}

{There have also been attempts to describe power law tail distribution, called $\kappa$-distributions, as some intermediate thermodnamic equilibrium in non-linear or turbulent interaction. Such attempts have so far been fairly artificial and academic being either based on a modification of Boltzmann's collision integral or on the proposition of some special form for the thermodynamic entropy depending on some free parameter like $\kappa$. Distributions of this kind fit fairly well observations. In how far they obey a thermodynamic state has, however, not been clarified from basic principles. It has also not been shown, how the free parameter $\kappa$ can be related to internal non-linear plasma processes on which it must necessarily depend if being real. In principle, of course, one could extend formally the derivation of an equivalent to the J\"uttner function also for such distributions. However, as long as there is no thermodynamic basis for them this remains a purely academic problem.} 

\subsection*{Various plasma quantities}
The Lorentz-invariant property of the phase-space volume element has consequences for the definition of a number of phase-space averages of plasma parameters. These come into play via the relativistic variation of the particle mass $m\to m\gamma(p)$ which depends on the single particle momentum. Hence, in presence of large particle numbers the collective averages of some quantities imply the phase-space integration over all contributing particles. 

\vspace{2mm}\noindent
\emph{Plasma frequency} 

\noindent The square of the plasma frequency of particles of rest mass $m_{0\alpha}$ and charge $q_\alpha$ is defined as $\omega^2_\alpha=q^2_\alpha N_\alpha/\epsilon_0m_\alpha$, where $m_\alpha=m_{0\alpha}\gamma(p)$ and $N$ is the particle density which follows from the zeroth moment of the relativistic distribution function. Under relativistic conditions with the single particles having proper Lorentz factors $\gamma(p)$, this becomes
\begin{equation}\label{plasmafreq}
\langle\omega_\alpha^2\rangle=\frac{q_\alpha^2}{\epsilon_0m_{0\alpha}}\int \frac{d^3xd^3p}{\gamma(p)}f_\alpha(\mathbf{p}) \to \omega_{0\alpha}^2\int \frac{d^3p}{\gamma(p)}f_\alpha(\mathbf{p})
\end{equation}
The extra Lorentz factor in the denominator of the integrand arises from the relativistic mass-dependence on the proper $\gamma(p)$. When introducing the rest-frame density $N^0_\alpha$ and assuming that the relativistic distribution is homogeneous in configuration space the integration over space can be performed leading to the second form in the above equation, where $\omega^2_{0\alpha}=q^2_\alpha N_{0\alpha}/\epsilon_0m_{0\alpha}$ is the square of the nonrelativistic plasma frequency.  The above expression takes into account that each particle has its own relativistic mass which depends on its proper momentum.

So far we assumed that the plasma in the observer's frame is at rest, in which case only the proper motions of the particles count. If there is a bulk flow with bulk momentum $\mathbf{p}_0$, then the un-primed frame can be considered as the rest frame of the plasma as a whole, and a new observer frame can be defined in which the plasma moves with bulk Lorentz factor $\Gamma(p_0)$. In this case the average density and mass transform like $N\to N\Gamma(p_0)$ and $\langle m_\alpha\rangle\to \langle m_\alpha\rangle\Gamma(p_0)$, where from Eq. (\ref{plasmafreq})
\begin{equation}
\langle m_\alpha\rangle=m_{0\alpha}\Big\{\int \frac{d^3p}{\gamma(p)}f_\alpha(\mathbf{p})\Big\}^{-1}
\end{equation}
yielding that the ratio of $N\Gamma(p_0)/\langle m_\alpha\rangle\Gamma(p_0)$ remains independent of $\Gamma(p_0)$. Thus, for bulk relativistic flows the bulk Lorentz factor $\Gamma(p_0)$ drops out of the plasma frequency, making it independent of any relativistic bulk flows.

\vspace{2mm}\noindent
\emph{Relativistic length scales}

\noindent The above definition of the plasma frequency leads to the relativistic redefinition of some lengths which depend on the plasma frequency. These are the Debye length and the inertial scales. The Debye length $\lambda_D$ does not depend on the particle mass. Because of this reason it remains invariant as long as no bulk flow is considered. In a medium flowing with bulk momentum $\mathbf{p}_0$ having a bulk Lorentz factor $\Gamma(p_0)$ its density dependence causes a weak dependence 
\begin{equation}
\lambda_D\to\lambda_{D0}/\sqrt{\Gamma(p_0)}
\end{equation}

As for the inertial scales $\lambda_\alpha=c/\omega_\alpha$, on the other hand, it is clear that the dependence of the plasma frequency makes them sensitive to the internal Lorentz factor as contained in the definition Eq. (\ref{plasmafreq}) of the plasma frequency. It is obtained by averaging the inverse inertial length taking into account that $f_\alpha$ is normalised
\begin{equation}
\left\langle\frac{c^2}{\lambda_\alpha^2}\right\rangle=\langle\omega_\alpha^2\rangle=\frac{c^2}{\langle\lambda_\alpha^2\rangle}
\end{equation}
which retains the functional form of the inertial length even in the relativistic case yielding for the root.mean-square of the inertial length of $\alpha$th particle species $\lambda_\alpha^\mathit{rel}=\sqrt{\langle\lambda_\alpha^2\rangle}=c/\sqrt{\langle\omega_\alpha^2\rangle}$, simply replacing the non-relativistic plasma frequency with its relativistic average. This form remains unchanged even in the case when the plasma moves at bulk Lorentz factor $\Gamma(p_0)$, for the reasons explained above.

\vspace{2mm}\noindent
\emph{Alfv\' en speed, Mach number, magnetisation parameter}

\noindent As a last quantity we consider the relativistic Alfv\'en velocity. The Alfv\'en speed is defined as $V_A=B/\sqrt{\mu_0mN}$ showing that it depends on the average particle mass density $mN$. Since $N$ is given as the zero moment of the distribution function, one generalises the mass density as
\begin{equation}
\langle m_\alpha N_\alpha\rangle= \int d^3\!\!p\ \gamma(p)\,f_\alpha(\mathbf{p})
\end{equation}
One may note that particles of high internal momentum contribute substantially to the relativistic mass density by becoming relativistically heavier. Moreover, if the plasma exhibits a bulk flow with Lorentz factor $\Gamma(p_0)$, then the mass density increases as $\Gamma^2$. 

Calculation of the relativistic Alfv\'en velocity can be done in analogy to calculating the inertial scale by taking $V_A^2$ and averaging its inverse over the phase space. This yields
\begin{equation}
V_A^\mathit{rel}=\sqrt{\langle V_A^2\rangle}=\frac{B}{\sqrt{\mu_0\langle m_\alpha N_\alpha\rangle}}
\end{equation}
Here the relativistic particles reduce the Alfv\'en speed due to their increasing mass effect. This also implies that the Mach number ${\cal M}=V/V_A$ will increase, if the plasma flows with bulk velocity $V=c\Gamma^{-1}\sqrt{\Gamma^2-1}$, in which case the Mach number becomes proportional to ${\cal M}^\mathit{rel}\approx{\cal M}_0 \sqrt{\Gamma^2-1}$, where ${\cal M}_0$ is the non-relativistic Mach number.

Finally, a comment on the magnetisation parameter (inverse plasma-$\beta$) is in place. This is defined as the ratio of the magnetic to kinetic energy densities
\begin{equation}
\sigma_m=\frac{B^2}{2\mu_0 NK}
\end{equation}
where $K=T+{\cal E}_\mathit{kin}$ is the sum of thermal and kinetic plasma energies, the latter playing a role only in plasma in bulk motion. A plasma not in bulk motion has thus constant magnetisation parameter $\sigma_m=$ const. However, when the plasma is in bulk relativistic motion, one has $N\to N\Gamma(p_0)$, and the magnetisation parameter decreases. Moreover, if the bulk energy is substantially larger than the thermal energy, as is the case in ultra-relativistic cool flows which are important in astrophysics, one has ${\cal E}_\mathit{kin}\propto \Gamma(p_0)\gg T$ and thus
\begin{equation}
\sigma_m=\frac{B^2}{2\mu_0 NK}\propto \frac{1}{\Gamma^2(p_0)}\ll 1
\end{equation}
which usually is very small and in this form is commonly used in ultra-relativistic astrophysical flows. 

Under this condition a flow can be considered as very weakly magnetised unless it becomes capable of generating its own strong equi-partition magnetic field. This is the case when the Weibel instability \citep{Weibel59,Fried59} takes over under some well defined conditions.

\conclusions
In summary, we have given here an argument for the Lorentz invariance of the phase-space volume element and constructed the corresponding relativistic distribution function. This has consequences for the calculation of some relevant plasma parameters in the relativistic domain. Application of such parameters can be found in the cyclotron-maser theory as for instance in the aurora or solar physics as well as in astrophysics, and to radiation belt problems where relativistic particles come into play. 

We have also shown that the \emph{isotropic thermal equilibrium distribution} function in a relativistic plasma {Eq. (\ref{reldist-1})} can be given analytically {[see Eq. \ref{lambda}]}; it has the form of a modified J\"uttner distribution. {Calculation of the anisotropic thermal distribution which has been argued is more correct than the isothermal \citep{israel1963,Kampen68,nakamura2009} in the relativistic case would be more involved and becomes possible only when the 4-vector of inverse temperature is properly taken into account. Since these questions have not yet been clarified even in basic relativistic thermodynamics, we have not invested any effort into the determination of the general relativistic distribution in that case.} The normalisation condition changes when the exact Lorentz invariance of the phase space volume is taken into account. This has consequences for the relativistic ideal gas laws, which have not been explored in this letter. {The interesting temperature-4-vector as well as anisotropic cases pose further hurdles and have not been considered here.}


\vspace{-1mm}
\begin{acknowledgements}
This research was part of an occasional Visiting Scientist Programme in 2006/2007 at ISSI, Bern. RT thankfully recognises the assistance of the ISSI librarians, Andrea Fischer and Irmela Schweizer. He highly appreciates the encouragement of Andr\'e Balogh, former Director at ISSI. The anonymous referee is thanked for his constructive comments.
\end{acknowledgements}
\vspace{-2mm}
\bibliographystyle{spbasic}      
\bibliography{Lorentz-1}   

\begin{thebibliography}{11}
\providecommand{\natexlab}[1]{#1}
\providecommand{\url}[1]{{#1}}
\providecommand{\urlprefix}{URL }
\expandafter\ifx\csname urlstyle\endcsname\relax
  \providecommand{\doi}[1]{DOI~\discretionary{}{}{}#1}\else
  \providecommand{\doi}{DOI~\discretionary{}{}{}\begingroup
  \urlstyle{rm}\Url}\fi
\providecommand{\eprint}[2][]{\url{#2}}

\bibitem[{{Chac{\'o}n-Acosta} et~al(2010){Chac{\'o}n-Acosta}, {Dagdug}, and
  {Morales-T{\'e}cotl}}]{dagdug2010}
{Chac{\'o}n-Acosta} G, {Dagdug} L, {Morales-T{\'e}cotl} HA (2010) {Manifestly
  covariant J{\"u}ttner distribution and equipartition theorem}. Phys Rev E
  81(2):021,126, \doi{10.1103/PhysRevE.81.021126}, \eprint{0910.1625}

\bibitem[{{Debbasch} et~al(2001){Debbasch}, {Rivet}, and {van
  Leeuwen}}]{debbasch2001}
{Debbasch} F, {Rivet} JP, {van Leeuwen} WA (2001) {Invariance of the
  relativistic one-particle distribution function}. Physica A: Statist Mech
  Appl 301:181--195, \doi{10.1016/S0378-4371(01)00359-4}, \eprint{0707.2499}

\bibitem[{{Dunkel} et~al(2007){Dunkel}, {Talkner}, and
  {H{\"a}nggi}}]{dunkel2007}
{Dunkel} J, {Talkner} P, {H{\"a}nggi} P (2007) {Relative entropy, Haar measures
  and relativistic canonical velocity distributions}. New J Physics 9:144--157,
  \doi{10.1088/1367-2630/9/5/144}, \eprint{arXiv:cond-mat/0610045}

\bibitem[{{Fried}(1959)}]{Fried59}
{Fried} BD (1959) {Mechanism for instability of transverse plasma waves}. Phys
  Fluids 2:337, \doi{10.1063/1.1705933}

\bibitem[{{Israel}(1963)}]{israel1963}
{Israel} W (1963) {Relativistic kinetic theory of a simple gas}. J Math Physics
  4:1163--1181, \doi{10.1063/1.1704047}

\bibitem[{{J{\"u}ttner}(1911)}]{juttner1911}
{J{\"u}ttner} F (1911) {Das Maxwellsche Gesetz der
  Geschwindig\-keits\-verteilung in der Relativtheorie}. Ann Physik (Leipzig)
  339:856--882, \doi{10.1002/andp.19113390503}

\bibitem[{{Landau} and {Lifshitz}(1975)}]{landau1975}
{Landau} LD, {Lifshitz} EM (1975) {The classical theory of fields}, vol~2.
  Pergamon Press, Oxford, UK

\bibitem[{{Nakamura}(2009)}]{nakamura2009}
{Nakamura} TK (2009) {Relativistic equilibrium distribution by relative entropy
  maximization}. Europhysics Letters 88:40,009,
  \doi{10.1209/0295-5075/88/40009}, \eprint{0909.2732}

\bibitem[{{van Kampen}(1968)}]{Kampen68}
{van Kampen} NG (1968) {Relativistic Thermodynamics of Moving Systems}. Phys
  Review 173:295--301, \doi{10.1103/PhysRev.173.295}

\bibitem[{{Weibel}(1959)}]{Weibel59}
{Weibel} ES (1959) {Spontaneously growing transverse waves in a plasma due to
  an anisotropic velocity distribution}. Phys Rev Lett 2:83--84,
  \doi{10.1103/PhysRevLett.2.83}

\bibitem[{{Weinberg}(1973)}]{weinberg1973}
{Weinberg} S (1973) {Gravitation and Cosmology: Principles and Applications of
  the General Theory of Relativity}. John Wiley \& Sons, Inc., New York,
  London, Sydney, Toronto

\end{thebibliography}
\vfill

\end{document}